\documentclass[journal]{IEEEtran}
\usepackage{amsmath,amsfonts}
\usepackage{algorithmic}
\usepackage{algorithm}
\usepackage{array}
\usepackage[caption=false,font=normalsize,labelfont=sf,textfont=sf]{subfig}
\usepackage{textcomp}
\usepackage{stfloats}
\usepackage{url}
\usepackage{verbatim}
\usepackage{graphicx}
\usepackage{cite}
\begin{document}

\title{Self-Supervised Learning for Audio-Based Emotion Recognition}

\author{Peranut~Nimitsurachat and Peter Washington}%



\maketitle

\begin{abstract}
Emotion recognition models using audio input data can enable the development of interactive systems with applications in mental healthcare, marketing, gaming, and social media analysis. While the field of affective computing using audio data is rich, a major barrier to achieve consistently high-performance models is the paucity of available training labels. Self-supervised learning (SSL) is a family of methods which can learn despite a scarcity of supervised labels by predicting properties of the data itself. To understand the utility of self-supervised learning for audio-based emotion recognition, we have applied self-supervised learning pre-training to the classification of emotions from  the CMU Multimodal Opinion Sentiment and Emotion Intensity (CMU-
MOSEI)'s acoustic modality. Unlike prior papers that have experimented with raw acoustic data, our technique has been applied to encoded acoustic data with 74 parameters of distinctive audio features at discrete timesteps. Our model is first pretrained to uncover the randomly-masked timestamps of the acoustic data. The pre-trained model is then fine-tuned using a small sample of annotated data. The performance of the final model is then evaluated via several evaluation metrics, including overall mean absolute error (MAE), mean absolute error (MAE) per emotion, overall 4-class accuracy, and 4-class accuracy per emotion. These metrics are compared against a baseline deep learning model with an identical backbone architecture. We find that self-supervised learning consistently improves the performance of the model across all metrics, especially when the number of annotated data points in the fine-tuning step is small. Furthermore, we quantify the behaviors of the self-supervised model and its convergence to baseline model as the amount of annotated data increases. This work shows the utility of self-supervised learning for affective computing, demonstrating that self-supervised learning is most useful when the number of training examples is small, and that the effect is most pronounced for emotions which are easier to classify such as happy, sad and anger. This work further demonstrates that self-supervised learning works when applied to embedded feature representations rather than the traditional approach of pre-training on the raw input space.
\end{abstract}


\section{Introduction}
\IEEEPARstart{E}{motion} classification has became increasingly used in a multitude of academic disciplines including digital psychiatry\cite{torous2021growing, washington2020data, washington2023review, pepa2021automatic}, autonomous vehicles\cite{izquierdo2018emotion, sini2020automatic}, and media analytics\cite{dai2015emotion, seng2017video}. By having sufficient labeled training data, researchers can improve the performance of emotion classifiers. A disproportionate amount of prior work in emotion recognition with video data places a stronger emphasis on the visual modality and often neglect the acoustic modality. Thus, deep learning models that classify emotions from audio input can potentially improve the performance of existing emotion classifiers.\par


A major challenge of training a performant emotion classifier for both the visual and acoustic modalities is the paucity of annotated data. Deep learning models are usually trained using a supervised-learning paradigm in which the model learns to map input--such as speech, facial movement, or facial expression--to the output or corresponding emotions. In order to attain high performance, the model usually requires large annotated training set; this type of dataset, however, tends to be scarce.
This lack of sufficient annotated data, therefore, makes training deep learning model to classify emotions accurately challenging. One solution is to manually annotate the unlabeled datasets with corresponding emotions; however, this task is very demanding and requires well-trained annotators. \par 

In this paper, we propose a self-supervision technique which can boost the accuracy of a deep learning model by taking advantage of additional pre-training steps on unannotated embedded audio data. We conduct our experiment on the CMU-MOSEI dataset \cite{zadeh2018multi}--one of the largest multimodal datasets to date.
We apply a similar methodology to the pre-training of large language models\cite{bert}, where we pre-train our model on acoustic data before fine-tuning on the small number of annotated emotion data. 
Pre-trained on the unlabeled data, we find that the model can learn useful internal representations of the embedded acoustic data which, in turn, can increase the model's accuracy in the fine-tuning stage even with a small number of annotated data. This work further demonstrates that self-supervised learning works when applied to the embedding space rather than the input space.
\begin{figure}[!t]
   \centering
   \includegraphics[width=3.2in, height=2.5in]{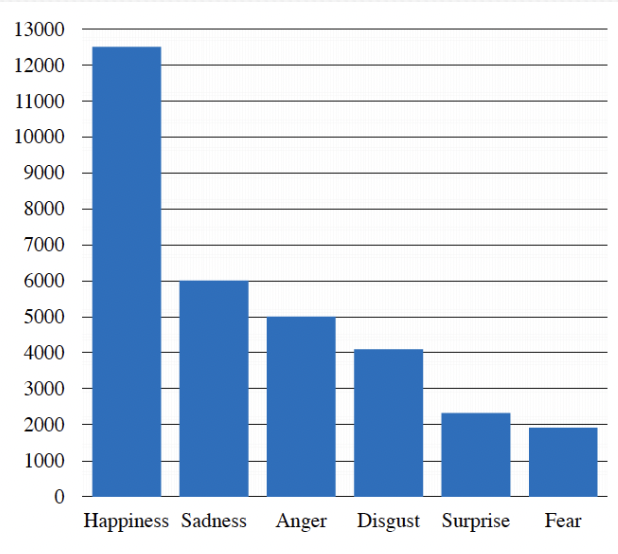}
   \caption{Distribution of emotions in CMU-MOSEI}
   \label{emo_dist}
\end{figure}

\begin{figure*}
   \centering
   \includegraphics[width=7in, height=4in]{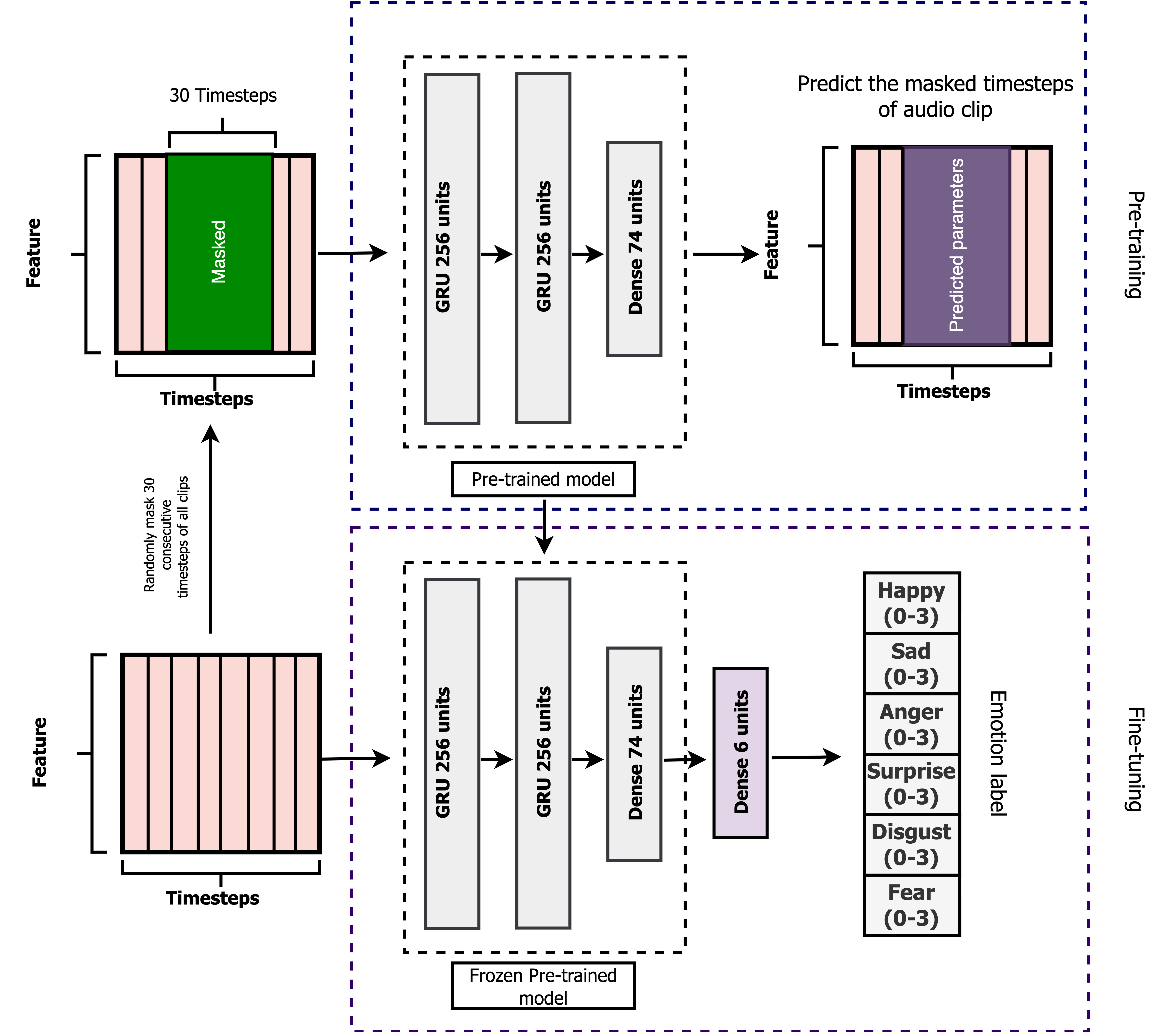}
   \caption{An overview of our proposed methodology for pre-training the deep learning model before transferring pre-trained model to the fine-tuning step}
   \label{overview}
\end{figure*}

\section{Related work}
Self-supervised learning has been applied mostly in natural language processing and computer vision. For instance,  the Bidirectional Encoder Representations from Transformers (BERT) \cite{bert} model pre-trains deep bi-directional representations on a large corpus through masked language modeling and next sentence prediction. The masked language modeling involves simply masking some percentage of the input tokens at random and predicting those masked tokens\cite{bert}. The pre-trained model is then transferred to a fine-tuning step where it is fine-tuned using annotated data. This pre-training technique helps improving the overall performance of the model especially when the number of labels is too small for deep learning.\par
Recent works have proposed to use similar self-supervised techniques on audio in raw format to perform several tasks such as speech recognition or emotion classification. In the case of emotion classification, Audio-only Self-Supervision (Odd) \cite{audio_self} jumbles 25 \% of the clips; two windows of a length of 15 \% of the selected clips are randomly selected and swapped. The encoder is then tasked to identify the elements in the input batch that have been swapped. During this pre-training step, the model will learn useful representation of the audio which can be beneficial in classifying discrete emotions in the fine-tuning step. Odd model explores this pre-training technique on several datasets, including CREMA-D\cite{crema}, Ravdess\cite{ravdess}, and IEMOCAP\cite{IEMOCAPIE}. All of these datasets store audio in raw format. Large language models like GPT-3 and GPT-4 \cite{gpt-3} also use similar strategy to predict the following word in the unlabeled text during the pre-training step.

Self-supervised techniques have also been applied to other models such as CPC (Contrast Predictive Coding)\cite{cpc} and APC (Autoregressive Predictive Coding)\cite{autoreg}. In the case of CPC (Contrast Predictive Coding), the authors propose a universal unsupervised learning approach to extract useful representations from high-dimensional data such as audio. This framework pre-trains the model by making predictions of the next token given the previous tokens. Specifically, it tries to learn representations that separate the target future frame from randomly sampled negative frames.  This pre-trained model is then fine-tuned and evaluated on phone prediction performance. CPC (Contrast Predictive Coding)\cite{cpc} is trained on the publicly available LibriSpeech dataset\cite{kaldids} and takes in audio inputs as 16KHz PCM audio waveform. Similary to CPC (Contrast Predictive Coding)\cite{cpc}, APC (Autoregressive Predictive Coding)\cite{autoreg} is also trying to predict the target future frames within the audio input. However, unlike CPC (Contrast Predictive Coding)\cite{cpc} that focuses on separating target frame from randomly sampled negative frame,  APC is only allowed to ignore information that is common across dataset. This APC framework\cite{autoreg} takes in audio data in the form of an 80-dimensional log Mel spectogram.\par

\begin{figure*}
   \centering
   \includegraphics[width=7in, height=4in]{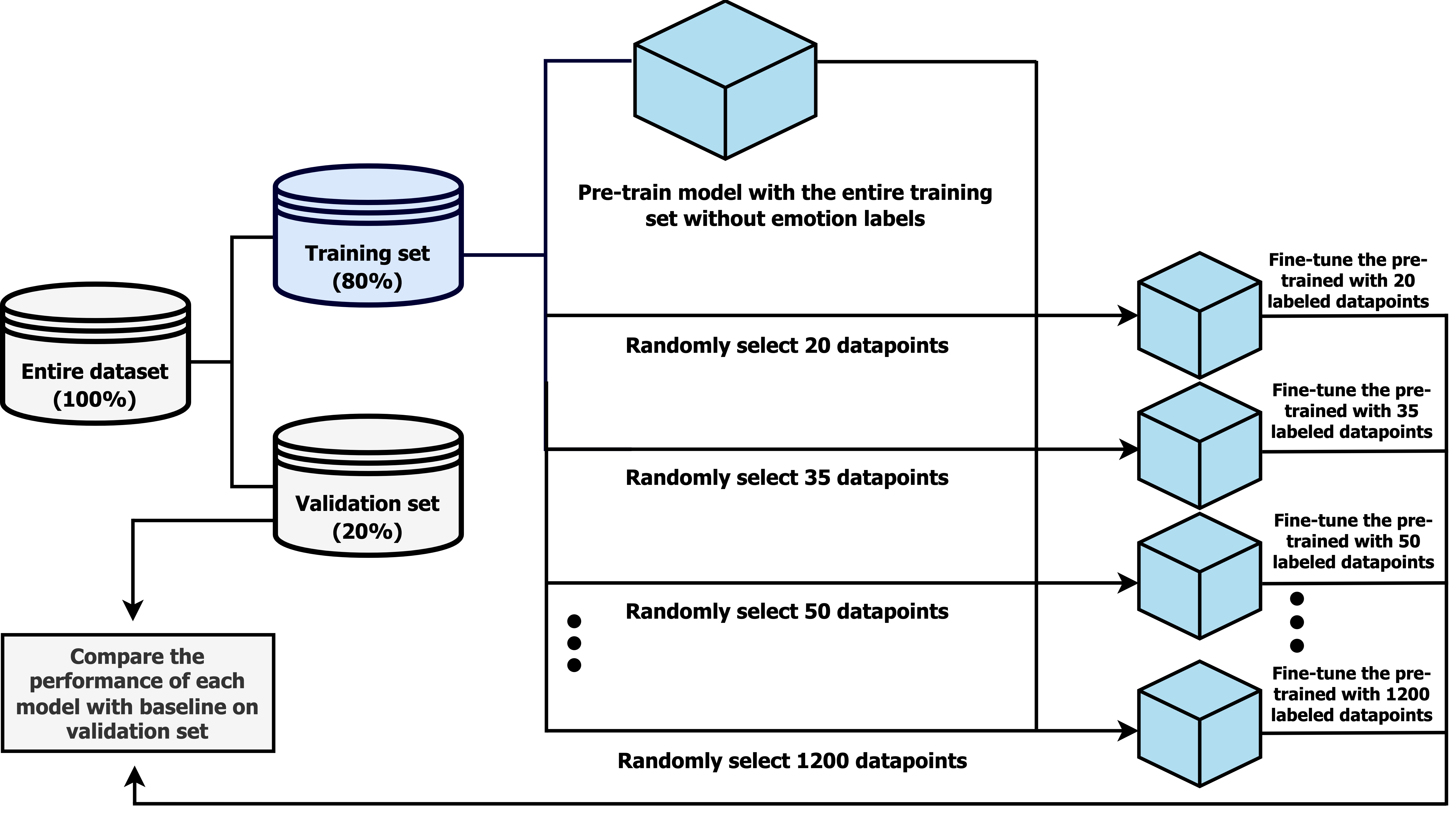}
   \caption{An overview of our experiment on how to pre-train and fine-tune the model with a variety of number of annotated datapoints and evaluate the performance of models in comparison to the baseline}
   \label{overview_experiment}
\end{figure*}

Another relevant work is Problem Agnostic Speech Encoder (PASE)\cite{pase}. This method pre-trains the model by predicting the seven extracted features--Waveform, Log power spectrum (LPS), Mel-frequency cepstral coefficients (MFCC), Prosody, Local info max (LIM),  Global info max (GIM), Sequence predicting coding (SPC)--of raw audio. This pre-trained model can be used for any downstream tasks by adding the linear classifier on top of frozen pre-trained layer. Finally, SeCoSt framework\cite{secost} also applied a teacher-student
self-supervised approach to audio by using the predictions of current iteration as labels for the next one. Similarly to previous works, the input features are  in form of logmel features. \par
As discussed above, most, if not all, of the previous works on self-supervised on audio apply this technique on the raw audio input. Our proposed method, however, will apply the masking technique to the encoded input data. Furthermore, we also monitor the performance of this technique for each of the six emotions in addition to the overall performance across all emotions. We are unaware of prior literature that characterizes this discrepancy.

\section{Methods}

\subsection{Dataset and features}
The experiment is conducted on CMU Multimodal Opinion Sentiment and Emotion Intensity (CMU-MOSEI)\cite{zadeh2018multi} which is considered to be one of the largest datasets of multimodal sentiment analysis and emotion recognition to date. This dataset contains more than 23,500 sentences utterance videos from 1,000 speakers on YouTube and has three modalities, including acoustic and visual. The dataset is also annotated with labels that indicate the level of six Ekman emotions\cite{ekmanemo} including happiness, sadness, anger, surprise, disgust ,and fear.  The labels are the array of six numbers; each one represents each of the intensity of each emotion $x$ on a scale of 0-3:  [0: no evidence
of $x$, 1: weakly $x$, 2: $x$, 3: highly $x$]\cite{bagher-zadeh-etal-2018-multimodal}. These annotations are assigned by 3 crowdsourced judges
from Amazon Mechanical Turk platform who received prior training and received 98\% or higher\cite{bagher-zadeh-etal-2018-multimodal} approval rate to assure high quality annotations. The distribution of emotions in this dataset is shown in 
Fig. \ref{emo_dist}.\par
Here, we work with the emotion labels and the acoustic modality of CMU-MOSEI. It is important to note that the acoustic data in CMU-MOSEI is not provided in a raw format; instead, each timestep of the input consists of 74 parameters that are extracted COVAREP software\cite{COVAREPA}. Therefore, we provide one of the first explorations of self-supervised learning on datasets which are encoded with an established feature embedding.


\begin{figure*}
   \centering
   \includegraphics[width=7in, height=3.26in]{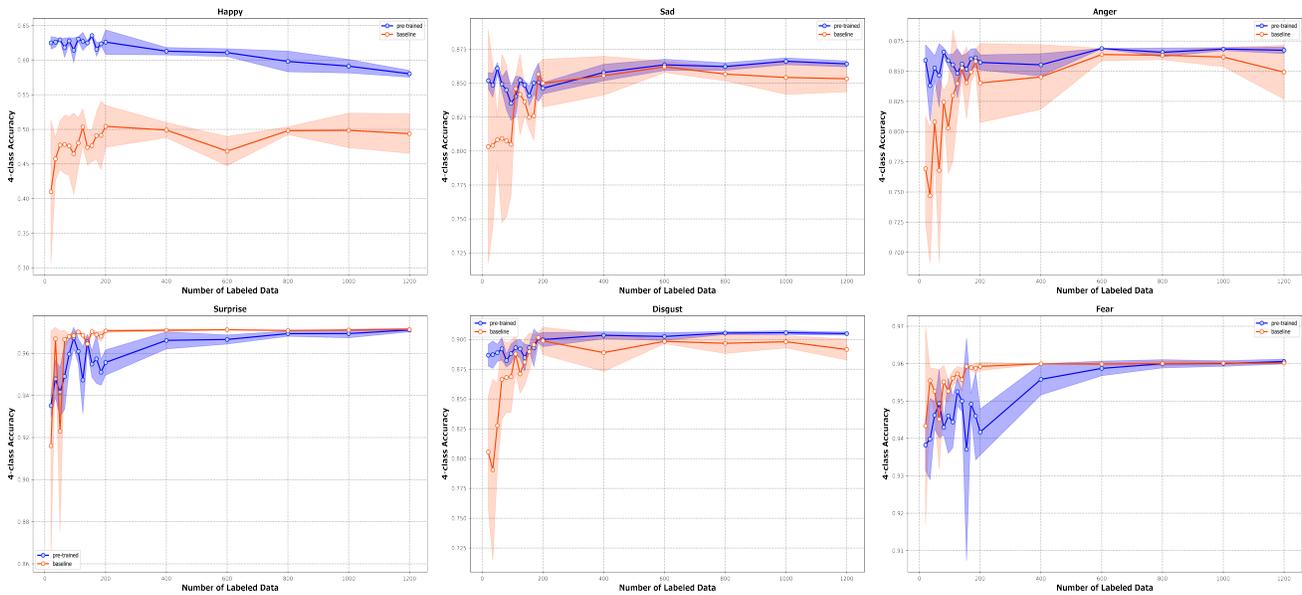}
   \caption{
   The 4-class accuracy for each of six emotions (Happy, Sad, Anger, Surprise, Disgust, Fear) along with the standard deviation calculated from three bootstrapped samples of the labeled data.}
   \label{accu_by_emo}
\end{figure*}

\begin{figure*}
   \centering
   \includegraphics[width=7in, height=3.26in]{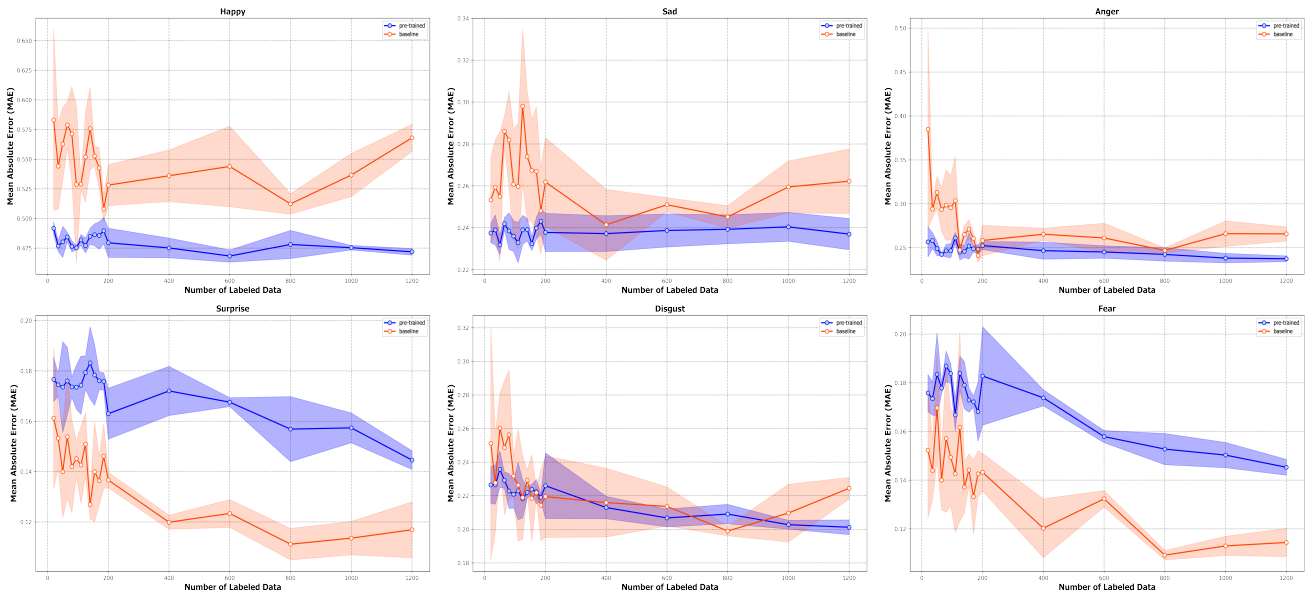}
   \caption{
   A plot showing the mean absolute error (MAE) for each of six emotions (Happy, Sad, Anger, Surprise, Disgust, Fear) along with the standard deviation calculated from three bootstrapped samples of the labeled data.}
   \label{mae_by_emo}
\end{figure*}
\subsection{Model details}
We pre-train a deep learning model to uncover the randomly masked timesteps within the audio clips. As mentioned in the previous section, each timestep of the audio input consists of 74 parameters extracted from COVAREP software\cite{COVAREPA}.\par 
Approximately 10\% of the audio input or 30 consecutive timesteps are randomly selected from every input; this is done by randomly selecting the valid starting timestep such that the following 30 timesteps are within the entire duration of input. As shown in Fig. \ref{overview}, all 74 parameters of the selected timesteps are masked or replaced with -30 which is completely outside of the standardized range of each parameter. \par
Our deep learning model, consisting of two layers of a 256-unit GRU followed by one 74-unit dense layer, is trained to predict the features of the original audio clip from the generated clip with masked features. By pre-training the model in this way, we encourage the pre-trained model to learn useful representations of the parameters of the audio clip that can be transferred to the fine-tuning step where the pre-trained model is trained with a small number of inputs annotated with an emotion label.\par
The pre-trained model is then transferred to the fine-tuning step. One 6-unit dense layer is added on top of the pre-trained model so that the new model can predict the emotion label, which is an array of 6 floating-point numbers, from the audio input. This 6-unit dense layer is then fine-tuned with a small fraction of audio input and corresponding emotion labels while all layers of pre-trained model are frozen. The performance of the final model is evaluated on the unseen validation set and compared with that of baseline model with a identical architecture.


\begin{figure}[!t]
   \centering
   \includegraphics[width=3.2in, height=2.5in]{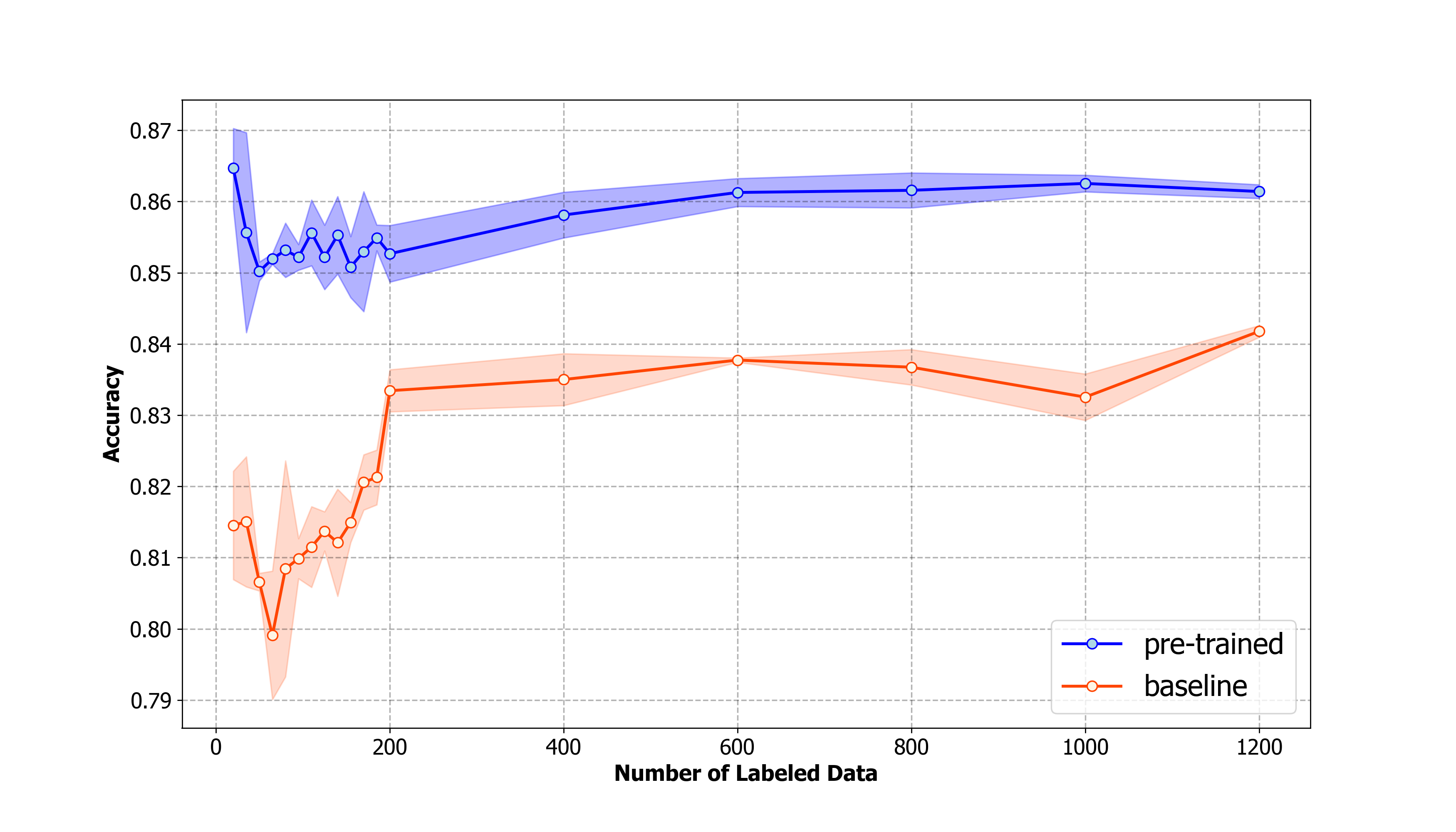}
   \caption{The 4-class overall accuracy across all six emotions (Happy, Sad, Anger, Surprise, Disgust, Fear) along with the standard deviation calculated from three bootstrapped samples of the labeled data.}
   \label{accu_over}
\end{figure}

\begin{figure}[!t]
   \centering
   \includegraphics[width=3.2in, height=2.5in]{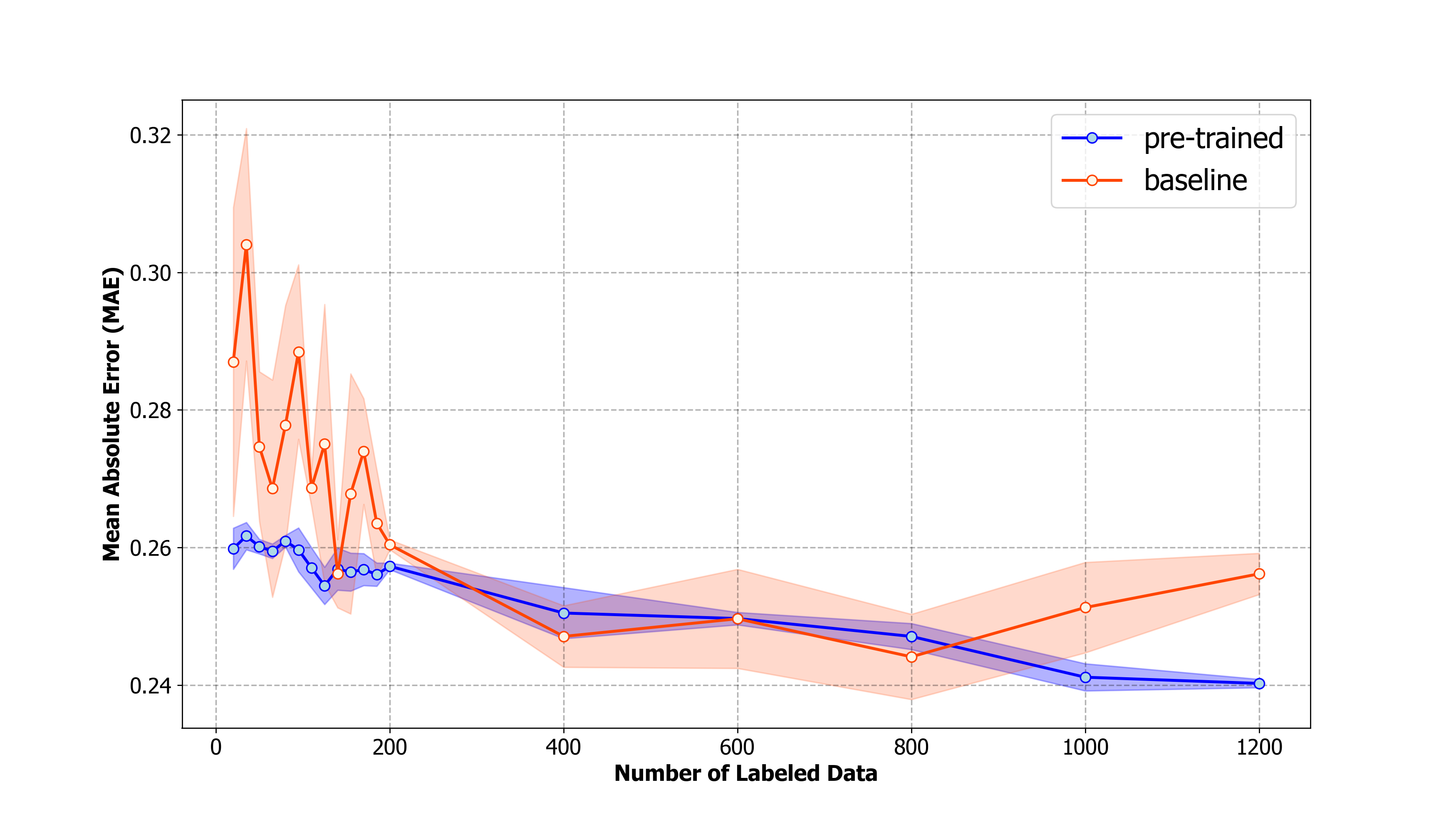}
   \caption{The MAE across all six emotions (Happy, Sad, Anger, Surprise, Disgust, Fear) along with the standard deviation calculated from three bootstrapped samples of the labeled data.}
   \label{mae_over}
\end{figure}

\subsection{Experiment}
We explore the performance of our deep learning model pre-trained with self-supervised learning in comparison to the baseline model with a identical architecture. The performance of the model is evaluated through the overall mean absolute error (MAE), mean absolute error (MAE) for each emotion, 4-class overall accuracy, and 4-class accuracy for each of six emotions.
\begin{equation}
\label{eqn:Rounding function}
r(x)=
    \begin{cases}
        \lfloor x \rfloor & \text{if } x - \lfloor x \rfloor < 0.5\\
        \lceil x \rceil & \text{if } x - \lfloor x \rfloor \geq 0.5
    \end{cases}
\end{equation}

Given the rounding function in Equation \ref{eqn:Rounding function}, the 4-class overall accuracy is defined as following. The metrics in Equation \ref{4-class} consist of six columns that represent the predicted and true values of six emotions of $n$ samples. 
\begin{equation}
\label{4-class}
\begin{aligned}
& f_{4-class} (
\begin{bmatrix}
x_{11}^{pred} & ... & x_{16}^{pred}\\
x_{21}^{pred} & ... & x_{26}^{pred}\\
...\\
x_{n1}^{pred} & ... & x_{n6}^{pred}\\
\end{bmatrix},
\begin{bmatrix}
x_{11}^{true} & ... & x_{16}^{true}\\
x_{21}^{true} & ... & x_{26}^{true}\\
...\\
x_{n1}^{true} & ... & x_{n6}^{true}\\
\end{bmatrix}) \\
& = \frac{ \sum_{k=1}^{n} \sum_{i=1}^{6} \textbf{1} [r(x_{ki}^{pred}) = r(x_{ki}^{true})]}{6n}
\end{aligned}
\end{equation}

The 4-class accuracy for each emotion is defined similarly. However, instead of considering all six columns of six emotions, only one column that corresponds to the emotion of interest is considered. \par

\begin{equation}
\label{4-class-accu}
\begin{aligned}
& f_{4-class}^{j} (
\begin{bmatrix}
x_{11}^{pred} & ... & x_{16}^{pred}\\
x_{21}^{pred} & ... & x_{26}^{pred}\\
...\\
x_{n1}^{pred} & ... & x_{n6}^{pred}\\
\end{bmatrix},
\begin{bmatrix}
x_{11}^{true} & ... & x_{16}^{true}\\
x_{21}^{true} & ... & x_{26}^{true}\\
...\\
x_{n1}^{true} & ... & x_{n6}^{true}\\
\end{bmatrix}) \\
& = \frac{ \sum_{k=1}^{n}  \textbf{1} [r(x_{kj}^{pred}) = r(x_{kj}^{true})]}{n}  \text{   when  }0\leq j \leq 6
\end{aligned}
\end{equation}

The overall mean absolute error (MAE) and mean absolute error (MAE) for each emotion use the standard definition of mean absolute error (MAE) on all six columns and one column of interest respectively.\par
As shown in Fig. \ref{overview_experiment}, the entire dataset is split into a training set (80\%) and validation set (20\%). The training set is used to train the baseline model and self-supervised models; the baseline model has the same architecture (two layers of 256-unit GRU and one 74-unit dense layer, one layer of 74-unit dense, and one layer of 6-unit dense) as the final pre-trained model shown in Fig.\ref{overview} but has no pre-training.
These two models are evaluated on the evaluation set using the four metrics mentioned above. The entire unlabeled training set is used to pre-train the model. We mask all 74 parameters of 30 randomly-selected consecutive timesteps and train a deep neural network model with two layers of a 256-unit GRUs and one 6-unit dense layer to uncover these masked timesteps. By predicting the masked timesteps of original audio input, the pre-trained model learns useful representation that can be transferred to the downstream task. It is imperative to note that we do not use the emotion labels of audio input at all in the pre-training step. This pre-trained model is then fine-tuned to emotion labels. As shown in Fig.\ref{overview_experiment}, we fine-tune our pre-trained model with different numbers of annotated data points in order to monitor the benefit of our pre-training technique when annotated data are scarce. In this experiment, we monitor the performance of the pre-trained and baseline models at 20, 35, 50, ..., 200 and 400, 600,..., 1200 data points. \par
A sample of data points are randomly selected from the training dataset along with their corresponding emotion labels; these annotated data points are then used to fine-tune the pre-trained model and train the baseline model for 30 epochs. The baseline and pre-trained models are then evaluated on the unseen evaluation set according to four performance metrics. We perform these three steps--randomly sampling, fine-tuning, and evaluating--three times per each number of data points in order to monitor the mean and variance of evaluation metrics.

\section{Results}
Fig.\ref{accu_over} shows the 4-class overall accuracy along with standard deviation of baseline and pre-trained model at each number of labeled data; the standard deviation in the figure is calculated from the accuracy of three iterations at each number of labeled data.
The 4-class overall accuracy of the pre-trained model is around 85-87 \% when there are a few number of labeled data (0-200). The accuracy gradually increases as the number of labeled increases. The accuracy of baseline model, on the other hand, starts at around 81-82 \% and increases rapidly as the number of labeled data approach 200. According to Fig. \ref{accu_over}, it is clear that the pre-trained model outperforms the baseline around 4-5 \% at small number of labeled data; the 4-class overall accuracy of both models starts to converge as the number of labeled data increase. It is also interesting to note that, for both baseline and pre-trained models, the standard deviations are relatively high with a few number of labeled data and becomes smaller once the number of labeled data increase.\par
Similarly, the overall mean absolute error (MAE) of the pre-trained model is around 0.26 at 20 labeled data and continues to shrink as the number of labeled data increases. The mean absolute error of the baseline model, however, is around 0.29-0.30 at 20 labeled data and shrinks as the number of labeled data increase. As shown in Fig. \ref{mae_over}, the performance of pre-trained model is better than the baseline especially at the small number of labeled data. Apart from higher accuracy and lower mean abosolute error, the standard deviations of these two metrics of pre-trained model are also lower than the baseline model. \par
As mentioned in previous section, we also monitor mean absolute error and 4-class accuracy for each emotion (happy, sad, anger, surprise, disgust, and fear). According to Fig.\ref{accu_by_emo}, the pre-trained model outperforms the baseline in accuracy of happy, sad, anger, and disgust by large margin. The largest gap between the accuracy of pre-trained and baseline models can be seen in happy and anger in which the pre-trained model outperforms by approximately 20\% and 10\% respectively. Our pre-trained technique, however, does not improve the accuracy of emotions like surprise and fear. Fig.\ref{accu_by_emo} shows that the baseline model performs slightly better than the pre-trained model. This can stem from the fact that these two emotions are relatively uncommon in this dataset according to the distribution in Fig \ref{emo_dist}. Thus, the pre-trained model might not be able to benefit from the large amount of unlabeled data in the pre-training step.  Furthermore, the accuracy of these two emotions are relatively high around 90\%; this can potentially make it difficult to improve beyond this threshold. \par
The mean absolute error for each emotion in Fig. \ref{mae_by_emo} also shows similar results. Firstly, we can see the large gap in the mean absolute error of baseline and pre-trained models in most emotions when the number of labels is small; this large gap then starts to shrink as the number of labels increase. The pre-trained model also outperforms the baseline with a large margin in traditionally "easier" emotions (e.g., happy, sad, and anger). The benefit of our pre-training technique, however, becomes marginal when coping with more nuanced expressions (e.g., surprise and fear). As discussed in the case of 4-class accuracy, the pre-trained model might not be able to benefit from the the large amount of unlabeled data in the pre-training step when classifying more nuanced expressions due to the scarcity of these emotions in the dataset. Furthermore, the mean absolute error of these nuance expressions are low at approximately 0.12-0.16 compared to the mean absolute error of traditionally "easier" emotions which can be as high as 0.4-0.5. Therefore, it might be difficult for our pre-trained model to improve beyond that.

\section{Discussion and Conclusion}

Consistent with prior self-supervised learning literature, we find that the benefits of self-supervised learning are observed when there are few supervised labels to learn from. Self-supervised learning provides marginal benefit when the number of labels is medium-to-large. This can be seen in both of 4-class accuracy and mean absolute error; the pre-trained model outperforms the baseline when the number of labels is small (0-200) and gradually converges to the baseline as the number of labels increase. Interestingly, we observe that the benefits of self-supervised learning are most pronounced for traditionally ``easier'' emotions (e.g., happy, sad, and anger) and more difficult for more nuanced expressions (e.g., surprise and fear). We are unaware of prior literature which characterizes this discrepancy. Furthermore, related works on self-supervised learning are usually applying this technique to the raw audio data. This work, on the other hand, applies self-supervised technique to the encoded audio data from COVAREP software\cite{COVAREPA}.

Although our pre-trained model can significantly outperform the baseline model in both evaluation metrics, there are some notable limitations to this work. Firstly, the baseline and pre-trained models are trained and evaluated for three iterations at each number of labels due to limited computing resources. We can obtain more stable results regarding the benefit of our pre-training technique by increasing the number of iterations at each number of labels. Furthermore, this work only applies self-supervised technique to audio modality. This could be another limitation since this technique can also work with other available modalities within CMU-MOSEI \cite{zadeh2018multi} or even cross-modality classification as discussed above.\par
Therefore, it would be interesting to investigate this pre-training technique on different encoded modalities within CMU-MOSEI \cite{zadeh2018multi} or datasets. Furthermore, we can also use the model pre-trained on audio modality to guide emotion recognition as well as any similar tasks on other available modalities (e.g., visual). We also want to explore alternative architectures of deep neural network (DNNs) that can potentially lead to better performance of our pre-trained model. This pre-training technique can also be applied to the raw audio data in which our results can be compared to the state-of-art methods from other related works.\par

On the other hand, the presented SSL methodology can be applied to datasets collected from video and audio data streams. While video-based emotion classifiers often use only visuals to make predictions, the addition of audio can bolster discriminative performance. These improved models can accelerate a variety of applications of affective computing. For example, \textit{SuperpowerGlass}\cite{glass, kline2019superpower, voss2016superpower, voss2019effect, daniels2018exploratory, kline2020superpower, washington2016wearable, daniels2018feasibility} is a wearable application that uses real-time computer vision to deliver real-time social cues to children with Autism Spectrum Disorder (ASD). The device uses the outward facing camera to read facial expressions of a conversation partner, and these expressions are classified into discrete Ekman emotions\cite{ekmanemo}. The device can provide children with real-time social cues reflecting the emotional expressions evoked by their conversation partners. \textit{Guess What?}, another digital therapeutic for children with ASD, also curates videos enriched for emotion expression\cite{guesswhat, kalantarian2019guess, kalantarian2018gamified, kalantarian2020performance, kalantarian2018mobile}. While these data have traditionally been used for purely visual prediction tasks\cite{washington2022improved, hou2021leveraging}, the addition of the audio modality combined with self-supervised learning can bolster performance to enable more personalized healthcare experiences. In general, self-supervised learning has the potential to learn the baseline temporal dynamics of data collected from a highly specialized domain distribution such as in ASD diagnostics. These `personalized' pre-trained models can be fine-tuned to traditionally complex target tasks like affect recognition. \par

\section*{Acknowledgments}
The authors would like to thank Kari Hanson for organizing Stanford ICME XPlore program. This program has played tremendous role in supporting this research from the beginning. Furthermore, the authors would like to thank ICME cloud computing cluster and Brian Tempero for providing computing resources Stanford ICME XPlore program and helping out with any technical difficulties.


\bibliographystyle{IEEEtran}
\bibliography{references}

\newpage

\section{Biography Section}
 
\vspace{11pt}

\vspace{-33pt}
\begin{IEEEbiography}[{\includegraphics[width=1in,height=1.25in,clip,keepaspectratio]{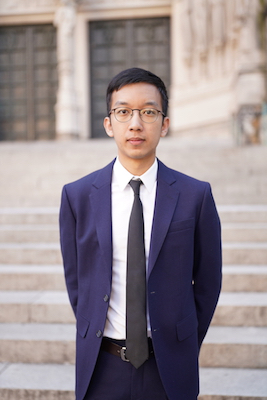}}]%
{Peranut Nimitsurachat}
received his bachelor's degree in Economics-Statistics and Computer Science from Columbia University in the City of New York in 2020. He is currently
a second year Master student in Institute of Computational and Mathematical Engineering (Data Science track) at Stanford University. His current research interests
include deep learning, emotion recognition, and reinforcement learning.
\end{IEEEbiography}

\begin{IEEEbiography}[{\includegraphics[width=1in,height=1.25in,clip,keepaspectratio]{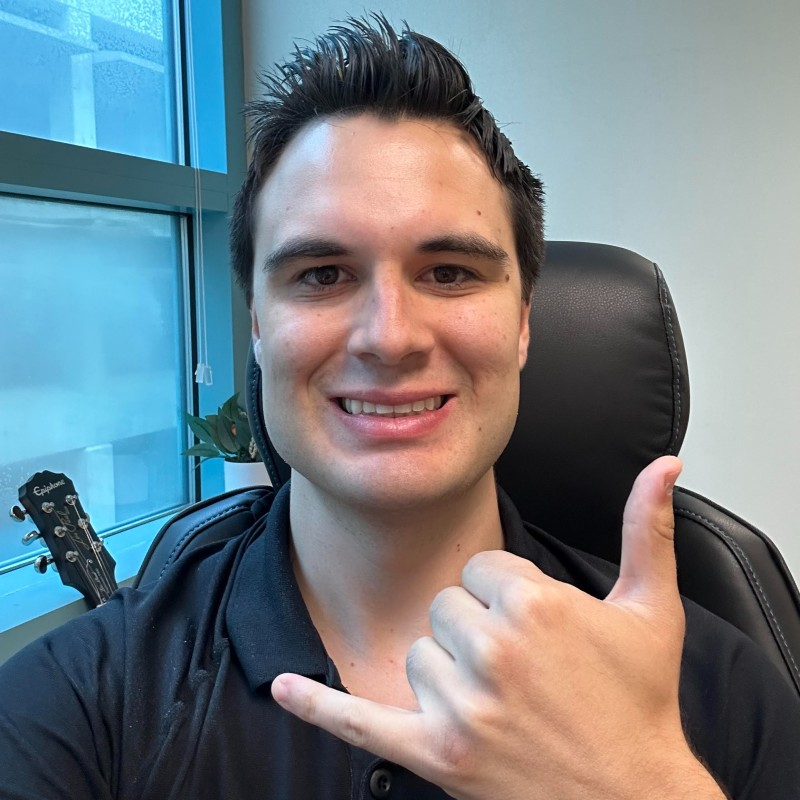}}]%
{Peter Washington}
is a tenure-track Assistant Professor in the Department of Information and Computer Sciences at the University of Hawaii at Manoa. His work spans digital health and machine learning innovations for healthcare with a particular focus on underserved populations in Hawaii. Prior to starting the Hawaii Digital Health Lab, he completed his PhD in Bioengineering from Stanford University, MS in Computer Science from Stanford University, and BA in Computer Science from Rice University.
\end{IEEEbiography}

\vspace{11pt}


\vfill

\end{document}